\documentclass[aps,11pt]{revtex4}
\usepackage[dvips]{graphicx}
\begin{document}

\title{Comment on Phys.Rev.Lett.'s paper: \\
"Low temperature Electron Spin Resonance \\
of the Kondo Ion in a heavy fermion metal YbRh$_{2}$Si$_{2}$"}
\author{R. J. Radwanski}
\homepage{http://www.css-physics.edu.pl} \email{sfradwan@cyf-kr.edu.pl}
\affiliation{Center of Solid State Physics,
S$^{nt}$ Filip 5, 31-150 Krakow, Poland,\\
@ Institute of Physics, Pedagogical University, 30-084 Krakow, Poland}
\author{Z. Ropka}
\affiliation{Center of Solid State Physics, S$^{nt}$ Filip 5, 31-150 Krakow, Poland}
\maketitle

The first successful Electron Spin Resonance (ESR) studies on single crystalline heavy fermion metallic compound
YbRh$_{2}$Si$_{2}$, recently reported by Sichelschmidt \textit{et al.} \cite{1}, has allowed for an unambiguous
observation of the localized $f$ states in this Kondo compound.

By this Comment \cite{2} we would like to congratulate Prof. F. Steglich and his group for the first successful
Electron Spin Resonance (ESR) on a heavy-fermion metallic compound at temperatures much below the Kondo temperature.
ESR has been usually performed on a diluted ionic system - they have managed to observe it on the paramagnetic ion
being the full part of a metallic solid. We fully agree with the interpretation of Ref. \cite{1} that the revealed
localized state is associated with the Kramers doublet of the Yb$^{3+}$ configuration, to be more specific - with the
4$f^{13}$ configuration of the Yb atoms.

Here we would like to report that the observed highly anisotropic $g$ tensor with $g_\bot$ = 3.561 and $g_\|$ = 0.17
can be perfectly described by crystal-field (CEF) interactions allowing for a small off-tetragonal orthorhombic
distortion. In the uniaxial CEF, whatever lower symmetry can be, there are only 4 Kramers doublets and in the
tetragonal symmetry the ground state can be only a state $\Gamma _{6}^{1}$ or $\Gamma _{7}^{1}$.

The perfect reproduction of the ESR results, $g_\bot$= 3.561 (with $g_L$ = 8/7 it corresponds to $J_\bot$ = $\pm 1.56$)
and $g_\|$ = 0.17 ($J_\|$ = $\pm 0.08$) is obtained for the $\Gamma _{7}^{1}$ ground state for parameters: $B_{2}^{0}$=
+14 K, $B_{4}^{0}$ = +60 mK, $B_{6}^{0}$ = -0.5 mK, $B_{4}^{4}$ = -2.30 K and $B_{6}^{4}$=-10 mK with a small local
orthorhombic distortion $B_{2}^{2}$ = +0.22 K. These parameters yield the ground doublet:

$\Gamma _{7}^{1}$ = 0.803 $|\pm 3/2>$ + 0.595 $|\mp 5/2>$ - 0.026 $|\mp 1/2>$ - 0.008 $|\pm 7/2>$

that is characterized by $J_\bot$ = $\pm 1.560$, $J_\|$ = $\pm 0.081$ and the quadrupolar operator value $Q_f$ =
$3J_{z}^{2}$ - J(J+1) of -4.7. The excited states are at 85 K ($\Gamma _{6}^{1}$ with $Q_f $ = -13.8), 485 K ($\Gamma
_{7}^{2}$) and 688 K ($\Gamma _{6}^{2}$). The admixture of last two small terms is an effect of the local orthorhombic
distortion.

In case of the ground state $\Gamma _{6}^{1}$ the perfect reproduction of the experimental ESR results can be attained
by:

$\Gamma _{6}^{1}$ = 0.944 $|\pm 1/2>$ + 0.322 $|\mp 7/2>$ - 0.052 $|\mp 3/2>$ - 0.046 $|\pm 5/2>$

that yields $J_\bot $ =$\pm 1.562$, $J_\|$ = $\pm 0.083$ and $Q_f$ of -11.2 This state can be obtained as the ground
state by tetragonal CEF interactions $B_{2}^{0}$ = +8 K, $B_{4}^{0}$ = -60 mK, $B_{6}^{0}$ = +6 mK, $B_{4}^{4}$ = -1.78
K, $B_{6}^{4}$ = - 5 mK with $B_{2}^{2}$ = +0.475 K, for instance. For these parameters the excited states are at 74 K
($\Gamma _{7}^{1}$ with $Q_f $ = -3.4), 430 K ($\Gamma _{6}^{2}$) and 437 K ($\Gamma _{7}^{2}$). This state is less
probable in comparison to the $\Gamma _{7}^{1}$ ground state as it gives too small anisotropy of the paramagnetic
susceptibility $\chi (T)$.

The electronic structure with the $\Gamma _{7}^{1}$ ground state is very plausible, though we do not think that it is
the final one. For it thermodynamical properties have to be more carefully analyzed, the best would be the direct
inelastic-neutron-scattering experiment. However, apart of the $g$ tensor the shown parameters reproduce the overall
temperature dependence of the paramagnetic susceptibility $\chi (T)$ and its huge anisotropy, presented in Fig. 1a of
Ref. \cite{3}, the preference for the magnetic ordering with moments perpendicular to the c axis, the magnetization
curve for external magnetic fields up to 60 T applied along the tetragonal c axis (Fig. 1b of Ref. \cite{3} - the
magnetization at 2 K and at 60 T amounts to 0.85 $\mu _{B}$). The derived electronic structure predicts a Schottky-type
contribution to the specific heat with a maximum of 3.62 J/Kmol at about T = 40 K, superimposed on the lattice heat,
and the anomalous temperature dependence of quadrupolar interactions. The excited doublet has much larger value of
$Q_f$ (in the absolute value) so with increasing temperature quadrupolar interactions, observed by means of the
Mossbauer spectroscopy, should pass a maximum as it was discussed in Ref. \cite{4}. Such the maximum is rather not
expected in case of the $\Gamma _{6}^{1}$ ground state. A maximum in $Q_f(T)$ has been observed in some Yb compounds
\cite{5}.

We do not think that the present set of CEF parameters is the final one. There is a plenty of sets that produce the
shown ground-state eigenfunction (the simplest can be obtained by multiplication by a constant positive value) but
surely the got set substantially confines the searching area of CEF parameters. We would like to express our big
surprise that, despite of a persistent advocating by last 12 years for the localized picture for the heavy-fermion
phenomena, so well-defined and so extremely thin atomic-scale energy levels exist in a metallic compound
YbRh$_{2}$Si$_{2}$ though we have observed extremely thin energy levels in conventional rare-earth intermetallics
\cite{6,7} and in a 3$d$ Mott insulator \cite{8}.

In short, we welcome with the great pleasure the ESR results of Prof. F. Steglich group on heavy fermion metal
YbRh$_{2}$Si$_{2}$ and we wish subsequent successful experiments on other Yb compounds. We fully agree with the F.
Steglich's interpretation claiming the existence of the localized $f$ electrons below Kondo temperature - this
interpretation concurs with our long lasting claim that the evaluation of CEF interactions is indispensable for the
physically adequate description of heavy-fermion compounds. We have derived CEF parameters of the tetragonal symmetry
with a small orthorhombic distortion that perfectly reproduce the ESR values ($g_\bot$ = 3.561 and $g_\|$ = 0.17) as
well as provide good reproduction of thermodynamical properties. The obtained parameters are surely not final as
searching for a consistent set of CEF parameters is like a large puzzle but the discussion of heavy-fermion compounds
within the same atomistic approach \cite{7,9} as conventional rare-earth compounds, where localized and band electrons
coexist, is very plausible from the unification point of view. Also from a methodological point of view - the
electronic structure and the importance of local distortions can be further experimentally verified. Finally, let
express our feeling that the paper of Ref. \cite{1} will be the turning point in the heavy-fermion subject.

\end{document}